# A Class of Random Sequences for Key Generation

Krishnamurthy Kirthi and Subhash Kak

**Abstract**:
This paper investigates randomness properties of sequences derived from Fibonacci and Gopala-Hemachandra sequences modulo m for use in key distribution applications. We show that for sequences modulo a prime a binary random sequence B(n) is obtained based on whether the period is p-1 (or a divisor) or 2p+2 (or a divisor). For the more general case of arbitrary m, we use the property if the period is a multiple of 8 or not. The sequences for prime modulo have much better autocorrelation properties. These are good candidates for key distribution since the generation process is not computationally complex.

**Introduction**

A good key sequence for cryptographic applications must have excellent randomness properties but also be easy to generate (Figure 1). Lacking this, keys can become the weak point of an otherwise strong cryptographic system. Here we propose the use of Fibonacci and the related Gopala-Hemachandra (GH) sequences [1]-[3] for this purpose. These sequences have applications in coding and cryptography that are well known. We consider the randomness properties of the residues of the Fibonacci sequence $F(n)$ = 0, 1, 1, 2, 3, 5, 8, …. and the related Gopala-Hemachandra sequence $GH_{a,b}(n)$ = a, b, a+b, a+2b, 2a+3b, … modulo $m$ (The sequence $GH_{a,b}$ will also be called (a,b)-GH). We do so by considering $m$ to be either a prime or as a composite number. Note that $GH_{a,b}(n) = GH_{a,b-1}(n) + F(n)$.

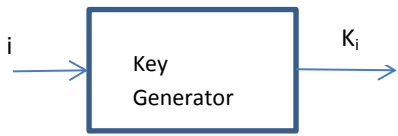

Figure 1. Key generation

The general period properties of such sequences are well known [4]-[6] but how these numbers indexed by *m* may be mapped into random sequences has not been investigated. We need good mappings to map these into binary sequence that have excellent autocorrelation properties. For different perspectives on randomness, that includes physical and algorithmic aspects, see [7]-[11].

Fibonacci and GH sequences are iterative. Another related iterative mapping is the algorithm to generate 1/p for prime p, the binary expansion a(n) of which is given by the formula [12]-[15] $a(n) = 2^n \bmod p \bmod 2$. Other related sequences of interest to the computer scientist and to the student of dynamical system theory include those obtained from general iterative maps [16]-[18].

In this paper, we summarize general properties of the periods of Fibonacci sequences mod *m*. Then, we present the mapping to transform these sequences into binary sequences that have excellent randomness properties.

**Periods of Fibonacci and GH sequences mod *m***

As mentioned before GH sequences can be written in terms of the Fibonacci sequence. Thus:

$GH_{2,1}(n) = F(n)+2F(n-1)$, for n > 1

As example, the sequence $GH_{2,1}(n)$ can be written as:

2, 1, 3, 4, 7, 11, … = 0, 1, 1, 2, 3, 5, 8, …

\+ 0, 0, 2, 2, 4, 6, 10, …

This indicates that the period of the GH sequence modulo *m* will be identical to that of the corresponding Fibonacci sequence.

Consider an F sequence to mod *m* series where *m* is a prime number. In this case, we can see that there can at most be $p^2-1$ pairs of consecutive residues in a period. For example, for m=p=3, the sequence will be 0, 1, 1, 2, 0, 2, 2, 1. This consists of the pairs 01, 02, 10, 11, 12, 20, 21, 22 which is all the possible pairs excepting 00, since that cannot be in such a sequence for it will lead to the next number being 0 that is impossible in a periodic residue sequence. Since $p^2-1$ is (p-1)(p+1), the period of the residue sequence will either be divisor of p-1 or p+1 (or equivalently of 2p+2).

The periods of Fibonacci sequence mod *m*, with *m* as a prime, has been shown to be either p-1 if p≡1 or 9 (mod 10) or 2p+2 if p≡3 or 7(mod 10). The periods of GH sequences likewise follow the same behavior.

The periods of generalized (a,b)-GH sequence mod *p* are [6]:

(i) (p-1) or a divisor thereof if the prime number *p* ends with 1 or 9.
(ii) (2p+2) or a divisor thereof if the prime number *p* ends with 3 or 7.
(iii) 20 for *p* =5.

The pertinent result for the period, N, of GH sequence for non-prime modulo *m* is:

N(m) ≤ 6m with equality iff $m = 2 \times 5^n$, for n=1, 2, 3, ….

This includes the value when the modulus is 5.

Thus, the periods can be grouped into 2 types: p-1 or 2p+2.

**Generating the sequence**

One can generate an arbitrary element of the F sequence by means of the following formula:

$$F(n) = \frac{1}{\sqrt{5}}(u^n - v^n) \bmod m$$

where

$$u = \frac{1+\sqrt{5}}{2}$$

and

$$v = \frac{1-\sqrt{5}}{2}$$

This result is easily proven by noting that $u^2 = u+1$ and $v^2 = v+1$. We get the sequence:

$$F(0) = 0; F(1) = (u-v)/\sqrt{5} = 1; F(2) = (u^2 - v^2)/\sqrt{5} = 1 \quad ; \quad F(3) = (u^3 - v^3)/\sqrt{5} = 2;$$
$$F(4) = (u^4 - v^4)/\sqrt{5} = 3; \text{ and so on.}$$

If n is a whole number, F(n) is generated and modulus of F(n) is calculated. Thus, F(n) mod *m* can be generated easily. On the other hand, obtaining n from F(n) mod *m* is difficult for large m, even if sufficient number of consecutive digits are available.

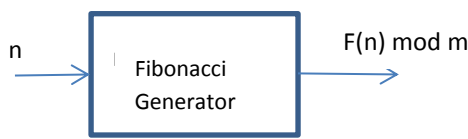

Figure 2. F(n) generator

**Mapping into random binary sequence**

As mentioned before, we propose to divide the sequence of periods for modulo p based on the property whether the period is a divisor of p-1 or 2p+2. We will include p=5 in the class p-1 since the period 20= 5x(5-1). We assign periods with multiples of (p-1) or divisor as binary value +1 and periods with multiples of (2p+2) or divisor as binary value -1.

Table 1 provides the first 25 prime numbers for easy reference.

Table 1: Binary mapping of prime periods

| Prime numbers | Periods | In terms of p | Binary value |
|---|---|---|---|
| 3 | 8 | 2p+2 | -1 |
| 5 | 20 | 5(p-1) | 1 |
| 7 | 16 | 2p+2 | -1 |
| 11 | 10 | p-1 | 1 |
| 13 | 28 | 2p+2 | -1 |
| 17 | 36 | 2p+2 | -1 |
| 19 | 18 | p-1 | 1 |
| 23 | 48 | 2p+2 | -1 |
| 29 | 14 | (p-1)/2 | 1 |
| 31 | 30 | p-1 | 1 |
| 37 | 76 | 2p+2 | -1 |
| 41 | 40 | p-1 | 1 |
| 43 | 88 | 2p+2 | -1 |
| 47 | 32 | (2p+2)/3 | -1 |
| 53 | 108 | 2p+2 | -1 |
| 59 | 58 | p-1 | 1 |
| 61 | 60 | p-1 | 1 |
| 67 | 136 | 2p+2 | -1 |
| 71 | 70 | p-1 | 1 |
| 73 | 148 | 2p+2 | -1 |
| 79 | 78 | p-1 | 1 |
| 83 | 168 | 2p+2 | -1 |
| 89 | 44 | (p-1)/2 | 1 |
| 97 | 196 | 2p+2 | -1 |
| 101 | 50 | (p-1)/2 | 1 |

We call the resulting binary sequence B(n). The first 20 bits of B(n) are -1,1,-1,1,-1,-1,1,-1,1,1,-1,1,-1,-1,-1,1,1,-1,1 and -1.

**Autocorrelation properties**

We first consider prime moduli and determine the autocorrelation properties of B(n) to determine how good they are from the point of view of randomness [15].

The autocorrelation function is calculated using the formula:

$$C(k) = \frac{1}{n}\sum_{j=0}^{n-1} B_j B_{j+k}$$

where $B_j$ and $B_{j+k}$ are the binary values of the sequence generated by the above process, and n is the length of the sequence.

Figures 3 and 4 present the normalized autocorrelation function of the B(n) sequence for 175 and 300 points.

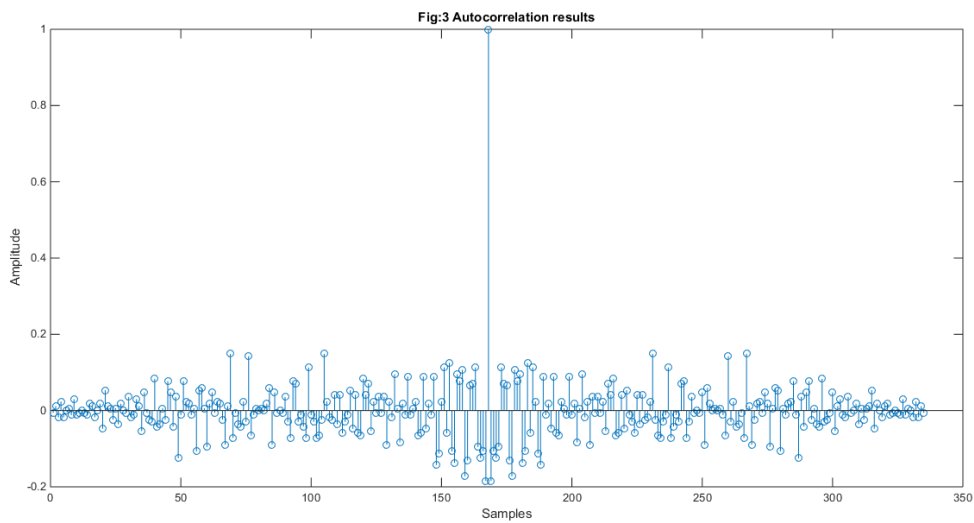

Figure 3. Autocorrelation of B(n) for sequence length 175

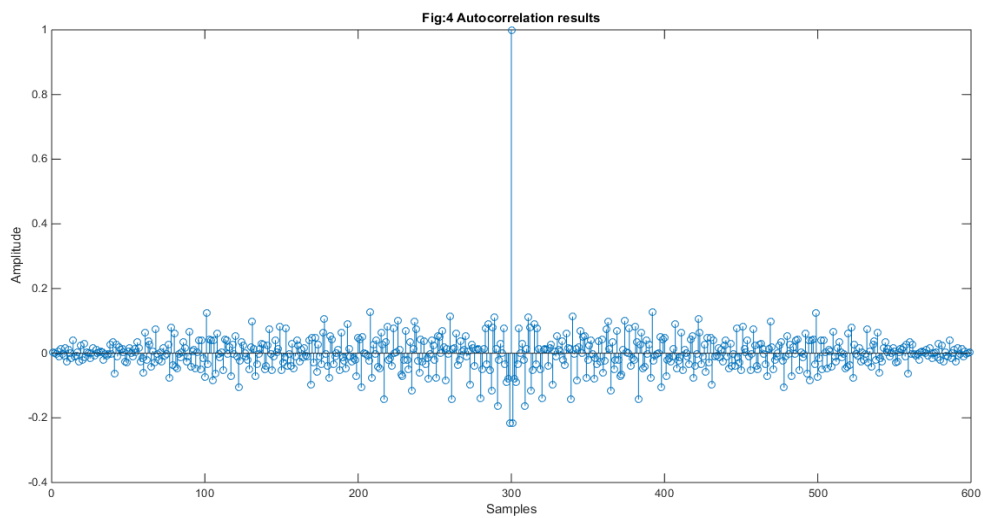

Figure 4. Autocorrelation of B(n) for sequence length 300

Looking at Figure 3 and Figure 4, their randomness is apparent from the effective two-valued character of the function.

Randomness may be calculated using the randomness measure, R(x), of a discrete sequence x by the expression below [8]:

$$R(x) = 1 - \frac{\sum_{k=1}^{n-1} |C(k)|}{n-1}$$

According to this measure a constant sequence will have the measure of 0 whereas a fully random sequence will have the measure of 1. The randomness measure values for Figure 3 and Figure 4 using the above formula is found to be 0.9516 and 0.9631.

General moduli *m*. The challenge is to find the property that helps map the period information into two classes that lead to a random binary sequence like B(n). For length of 300, we found the property whether the period is a multiple of 8 to effectively put the period values into two classes. But the randomness measure of such a sequence was much inferior compared to that of Figures 3 and 4 with a value of 0.8988.

**Conclusion**

We have shown that for Fibonacci and GH sequences modulo a prime a binary random sequence B(n) is obtained based on whether the period is p-1 (or a divisor) or 2p+2 (or a divisor). For the more general case of arbitrary m, we used the property if the period is a multiple of 8 or not. The sequences for prime modulo have excellent autocorrelation properties. These are good candidates for key generation since the generation process is not computationally complex.

**References**

1. J. Wu, Extended Fibonacci cubes. IEEE Trans. on Parallel and Distributed Systems 8, 1203-1210, 1997.
2. J.H. Thomas, Variations on the Fibonacci universal code. arXiv:cs/0701085 (2007)
3. M. Basu and B. Prasad, Long range variations on the Fibonacci universal code. Journal of Number Theory 130: 1925-1931 (2010)
4. A. Nalli and C. Ozyilmaz, The third order variations on the Fibonacci universal code. Journal of Number Theory 149: 15-32 (2015)
5. S. Gupta, P. Rockstroh, and F.E. Su, Splitting fields and periods of Fibonacci sequences modulo primes, Math. Mag. 85: 130–135 (2012)
6. M. Renault, The period, rank, and order of the (a,b)-Fibonacci sequence mod m. Math. Mag. 86: 372-380 (2013)
7. A Kolmogorov, Three approaches to the quantitative definition of information. Problems of Information Transmission. 1:1-17 (1965)

8. S. Kak, Classification of random binary sequences using Walsh-Fourier analysis. IEEE Trans. on Electromagnetic Compatibility, EMC-13: 74-77 (1971)
9. S Kak, Information, physics and computation. Found. of Phys. 26: 127-137 (1996)
10. R. Landauer, The physical nature of information. Physics Letters A 217: 188-193 (1996)
11. S Kak, Quantum information and entropy. Int. Journal of Theo. Phys. 46: 860-876 (2007)
12. S. Kak and A. Chatterjee, On decimal sequences. IEEE Trans. on Information Theory IT-27: 647 – 652 (1981)
13. S. Kak, Encryption and error-correction coding using D sequences. IEEE Trans. on Computers C-34: 803-809 (1985)
14. N. Mandhani and S. Kak, Watermarking using decimal sequences. Cryptologia 29: 50-58 (2005)
15. S.B. Thippireddy, Binary random sequences obtained from decimal sequences. arXiv preprint arXiv:0809.0676 (2008)
16. S. Kak, Feedback neural networks: new characteristics and a generalization. Circuits, Systems, and Signal Processing 12: 263-278 (1993)
17. N.S. Korripati, Random sequences generated by recursive maps. https://subhask.okstate.edu/sites/default/files/nikhil11.pdf
18. S. Kak, Goldbach partitions and sequences. Resonance 19: 1028-1037 (2014)